\documentclass{article}
\begin{document}

\begin{title}
\centerline {\large \bf Problem of water vapor absorption continuum 
\linebreak in atmospheric windows. Return of dimer hypothesis.}
\end{title}

\vspace{7 pt}
\centerline{\sl V.A.Kuz'menko}
\vspace{7 pt}
\centerline{\small \it Troitsk Institute for Innovation and Fusion Research,}
\centerline{\small \it Troitsk, Moscow region, 142190, Russian Federation.}
 
\vspace{5 pt}
\begin{abstract}

	 Two alternative hypotheses try to explain the origin of the 
continuum. The present communication gives new argument in favor of the 
dimer hypothesis. This argument is based on the existence of the wide 
component of line in absorption spectrum of polyatomic molecules.
\end{abstract}
\vspace{12 pt}

	The continuum of water vapor absorption in the region of atmospheric
windows (8--14  and  3--5 $\mu m$) was found in 1938 [1]. The first 
explanation of its nature was that continuum is the sum of far wings of 
collisionally broadened lines of rotational and vibrational-rotational 
transitions in water molecules.
 
	Afterwards an anomalously strong square dependence of absorption on
 water vapor pressure was observed along with its exponential negative 
temperature dependence [2,3]. The slope of the exponent is 5 kcal/mole and is
 thus equal to the binding energy of molecules in a dimer of water. These are
 strong arguments in favor of the dimer hypothesis which assumes that the 
continuum is caused by the presence of water dimers in the gas. But these are 
indirect proofs. In succeeding years no new facts in favor of the dimer 
hypothesis were found. But it was definitely determined that no intensive 
absorption bands of water dimer molecules are present in the region of 
atmospheric windows. This is considered to be the main argument against the 
dimer hypothesis. In addition, the shape of the continuum corresponds better 
to far wings of the lines than to some local bands of molecules absorption. 
As a result, in succeeding years most scientists again gave preference to 
the monomer hypothesis [4].

	Thus, it became a deadlock situation. There existed explicit indirect 
indications of the dimer hypothesis. Direct spectroscopic studies seem 
contradict it and show the validity of the monomer hypothesis, which fails 
to explain satisfactorily the temperature dependence.

	The present communication is aimed at drawing the attention of 
scientists to a new argument in favor of the dimer hypothesis. A rather 
elegant way to end the deadlock now exists: the continuum is 
actually formed by far wings of molecule absorption lines. But these are 
mainly the wings of absorption lines of water dimer molecules. The lines of 
dimer molecules  have  intensive  Lorentzian  wings,  whose absorption cross 
section can be several orders of magnitude  higher than for monomers. 

	Such unconventional supposition has come from another field of 
scientific investigations, namely, from those of IR multiple-photon 
excitation of molecules. The existence of broad intensive line wings 
(or wide component of line) in the absorption spectrum of polyatomic 
molecules was shown in [5]. There are many proofs of the existence of 
these wings. For example, for number of polyatomic molecules the far 
Lorentzian wings of absorption bands in the gaseous phase were founded. 
Their intensity is in good agreement with the data on the saturation degree 
of a linear spectrum of molecules absorption by a pulse $CO_2$-laser 
radiation at low gas pressure. On the basis of these data the relative 
integral intensity of line wings can easily be evaluated. It is 
$\sim 0.6\%$ for $SF_6$ and $SiF_4$ molecules, $11\%$ for $CH_3SiF_3$ 
molecules, $\sim 90\%$ for $(CF_3)_2O$ and $(CF_3)_2CO$ molecules and $>90\%$ 
for $C_6H_5SiF_3$ molecules. 

	One can see that the intensity of the line wings grows rapidly with 
the number of atoms in a molecule. Accordingly, it can be expected that in 
molecule dimers the intensity of line wings will be considerably higher than 
for monomers. No data are available yet on the intensity of line wings in 
water dimers. But for ethylene dimers the published experimental results 
[6--8] allow such evaluation to do. Nearly 80\% of the integral intensity of 
the absorption line of ethylene dimers is in it wings.

	Thus, the problem of the continuum of water vapor absorption will be 
totally clarified when the width and the intensity of the line wings of 
rotational and vibrational-rotational transitions of water dimer  
molecules are measured [9]. For their detection more powerful lasers (CO and 
HF-lasers) should be used than that used in the work [10].


\vspace{5 pt}
\begin{thebibliography}{99}
\bibitem{1}W.M.Elsasser, Astrophys. J. {\bf 87},497 (1938). 
\bibitem{2}A.A.Victorova and S.H.Zhevakin, Sov.Phys.Dokl. {\bf 11}, 
1059 (1967). 
\bibitem{3}P.Varanasi, S.Chou and S.S.Penner, 
J.Quant.Spectrosc.Radiat.Transfer {\bf 8}, 1537 (1968). 
\bibitem{4}Q.Ma and R.H.Tipping, J.Chem.Phys. {\bf 93}, 7066 (1990). 
\bibitem{5}V.A.Kuz'menko, E-print, physics/0204003
\bibitem{6}M.A.Hoffbauer, K.Liu, C.F.Giese and W.R.Gentry, J.Chem.Phys.
{\bf 78}, 5567 (1983). 
\bibitem{7}B.Heijmen, C.Liedenbaum, S.Stolte and J.Reuss, Z.Phys.D  
{\bf 6}, 199 (1987) 
\bibitem{8}U.Back, F.Huisken, Ch.Lauenstein, H.Meyer and R.Sroka, J.Chem.Phys. 
{\bf 87}, 6276 (1987).
\bibitem{9}P.Varanasi, J.Quant.Spectrosc.Radiat.Transfer {\bf 40}, 169 (1988).
\bibitem{10}Z.S.Huang and R.E.Miller, J.Chem.Phys. {\bf 91}, 6613 (1989). 
\end{thebibliography}
\end{document}